%% file: HIN-12-007_temp.tex
\pdfoutput=1

\documentclass[11pt,twoside,a4paper,cmspaper,final,collab]{cms-tdr}
\def\svnVersion{272952}\def\cmsCernNo{2013-037}\def\cmsCernDate{\today}\def\cmsMessage{Published in Physical Review Letters as \href{http://dx.doi.org/10.1103/PhysRevLett.113.262301}{\doi{10.1103/PhysRevLett.113.262301.}}}

\begin{document}\cmsNoteHeader{HIN-12-007}

\hyphenation{had-ron-i-za-tion}
\hyphenation{cal-or-i-me-ter}
\hyphenation{de-vices}
\RCS$Revision: 266143 $
\RCS$HeadURL: svn+ssh://svn.cern.ch/reps/tdr2/papers/HIN-12-007/trunk/HIN-12-007.tex $
\RCS$Id: HIN-12-007.tex 266143 2014-11-02 21:52:28Z tdahms $
\newlength\cmsFigWidth
\ifthenelse{\boolean{cms@external}}{\setlength\cmsFigWidth{0.98\columnwidth}}{\setlength\cmsFigWidth{0.6\textwidth}}
\ifthenelse{\boolean{cms@external}}{\providecommand{\cmsLeft}{top}}{\providecommand{\cmsLeft}{left}}
\ifthenelse{\boolean{cms@external}}{\providecommand{\cmsRight}{bottom}}{\providecommand{\cmsRight}{right}}
\ifthenelse{\boolean{cms@external}}{\providecommand{\CL}{C.L.\xspace}}{\providecommand{\CL}{CL\xspace}}

\newcommand{\psiP}{\ensuremath{\psi\text{(2S)}}\xspace}
\newcommand{\raa}{\ensuremath{R_\mathrm{AA}}\xspace}
\newcommand{\pp}{{\ensuremath{\Pp\Pp}}\xspace}
\newcommand{\PbPb}{\ensuremath{\text{PbPb}}\xspace}
\newcommand{\pPb}{\ensuremath{\Pp\text{Pb}}\xspace}
\newcommand{\dAu}{\ensuremath{\text{dAu}}\xspace}
\newcommand{\doubleRatio}{\ensuremath{\left.(N_{\psiP}/N_{\JPsi})_{\PbPb}/(N_{\psiP}/N_{\JPsi})_{\pp}\right.\xspace}}
\newcommand{\npart}{\ensuremath{N_{\text{part}}}\xspace}
\newcommand{\mumu}{\ensuremath{\Pgmp\Pgmm}\xspace}
\providecommand{\HYDJET}{{\textsc{hydjet}}\xspace}

\newcommand{\sqrts}{\ensuremath{\sqrt{s}}\xspace}
\newcommand{\sqrtsnn}{\ensuremath{\sqrt{s_{_{\text{NN}}}}}\xspace}

\cmsNoteHeader{HIN-12-007} % This is over-written in the CMS environment: useful as preprint no. for export versions
\title{\texorpdfstring{Measurement of prompt $\psiP$ to $\JPsi$  yield ratios
in \PbPb and \pp collisions at $\sqrtsnn=2.76\TeV$}{Measurement of prompt psi(2S) to J/psi yield ratios in PbPb and pp collisions at sqrt(s[NN]) = 2.76 TeV}}

\date{\today}

\abstract{ The ratio between the prompt \psiP and
  $\JPsi$ yields, reconstructed via their
  decays into $\Pgmp\Pgmm$, is measured in \PbPb and \pp collisions at
  $\sqrtsnn = 2.76$\TeV. The analysis is based on \PbPb and \pp data
  samples collected by CMS at the Large Hadron Collider, corresponding to integrated luminosities
  of 150\mubinv and 5.4\pbinv, respectively. The double ratio of
  measured yields, $(N_{\psiP} /
  N_{\JPsi})_{\PbPb} /
  (N_{\psiP} /
  N_{\JPsi})_{\pp}$, is computed in
  three PbPb collision centrality bins and two kinematic ranges: one at
  midrapidity, $\abs{y}<1.6$, covering the transverse momentum range $6.5<\pt<30\GeVc$, and the other at forward rapidity, $1.6<\abs{y}<2.4$,
  extending to lower \pt values, $3<\pt<30\GeVc$. The
  centrality-integrated double ratio changes from
  $0.45\pm0.13\stat\pm0.07\syst$ in the first range to
  $1.67\pm0.34\stat\pm0.27\syst$ in the second. This
  difference is most pronounced in the most central collisions.}

\hypersetup{%
  pdfauthor={CMS Collaboration},%,
  pdftitle={Measurement of prompt psi(2S) to J/psi yield ratios in PbPb and pp collisions at sqrt(s[NN]) = 2.76 TeV},%
  pdfsubject={CMS},%
  pdfkeywords={CMS, physics, heavy ions, dimuons, quarkonia}}

\maketitle %maketitle comes after all the front information has been supplied

The goal of the study of ultrarelativistic heavy-ion collisions is to create and
characterize the quark-gluon plasma (QGP), a medium where quarks and gluons are
no longer confined in hadrons~\cite{Karsch:2003jg}. Charmonia should dissociate
when the Debye screening radius of this medium, which decreases with increasing
QGP temperature, becomes smaller than the binding radius of the charmonium
state~\cite{Matsui:1986dk}. Since the \psiP meson is less bound than the \JPsi,
it should melt at lower temperatures~\cite{Karsch:2005nk}, an idea consistent
with charmonium suppression measurements made at the
CERN~SPS~\cite{Baglin:1994ui,Alessandro:2006ju}.
At the CERN LHC, a suppression of \JPsi mesons in \PbPb collisions at a
center-of-mass energy per nucleon-nucleon pair of $\sqrtsnn=2.76\TeV$ was
observed by CMS~\cite{Chatrchyan:2012np} and
ALICE~\cite{Abelev:2012rv,Abelev:2013ila} via the nuclear modification factor,
\raa, the ratio of nucleus-nucleus and \pp charmonium production yields
normalized by the number of inelastic nucleon-nucleon collisions. The
suppression increases with transverse momentum (\pt), exhibiting a strong
centrality dependence at high \pt, but almost no dependence when integrated over
\pt. Related results at the SPS and BNL RHIC are presented in
Ref.~\cite{Brambilla:2010cs} and references therein.

Given the large number ($\mathcal{O}(100)$) of charm quarks produced per central
\PbPb collision at $\sqrtsnn=2.76\TeV$~\cite{Zhao:2011cv}, charmonia may also be
produced at the hadronization stage, through the combination of
initially-uncorrelated charm and anticharm
quarks~\cite{Thews:2000rj,Andronic:2006ky}. This mechanism should contribute
mostly at low \pt~\cite{Zhao:2011cv}.
Charmonium production is also affected by ``cold nuclear matter''
effects~\cite{Vogt:2010aa,Zhao:2011cv,Arleo:2013xza}, such as nuclear modifications of
the parton distribution functions.
Recently, ALICE~\cite{Abelev:2013yxa} and LHCb~\cite{Aaij:2013zxa} observed
\JPsi suppression in \pPb collisions, while PHENIX~\cite{Adare:2013ezl} and
ALICE~\cite{Abelev:2014zpa} reported that the \psiP is more strongly suppressed
than the \JPsi in \dAu and \pPb collisions, complementing analogous observations
made by fixed-target experiments~\cite{Lourenco:2008sk}.

This Letter presents a measurement of the prompt \psiP and \JPsi yields
(excluding production from decays of \cPqb\ hadrons) in \PbPb and \pp collisions
at $\sqrtsnn = 2.76\TeV$, using event samples collected by CMS with integrated
luminosities of $150\mubinv$ and $5.4\pbinv$, respectively. Following related
studies of the bottomonium
family~\cite{Chatrchyan:2011pe,Chatrchyan:2012lxa,Chatrchyan:2013nza}, the
results are reported as a double ratio, \doubleRatio, so that efficiency and
acceptance corrections cancel to a large extent, reducing the systematic
uncertainties. Using a previous measurement of
$\raa(\JPsi)$~\cite{Chatrchyan:2012np}, a first measurement of $\raa(\psiP)$ is
derived.

The central feature of CMS is a superconducting solenoid of 6\unit{m} internal
diameter, providing a magnetic field of 3.8\unit{T}. Within the field volume are the
silicon tracker, the crystal electromagnetic calorimeter, and the
brass/scintillator hadron calorimeter. The silicon pixel and strip tracker
measures charged-particle trajectories in the pseudorapidity range $\abs{\eta}<2.5$.
Muons are detected in the interval $\abs{\eta}< 2.4$ by gas-ionization detectors
made of three technologies: drift tubes, cathode strip chambers, and
resistive-plate chambers. The muon \pt is measured with a resolution between 1
and 2\% for a typical muon in this analysis. Two steel/quartz-fibre
Che\-ren\-kov forward hadron (HF) calorimeters cover the range $2.9 < \abs{\eta}<
5.2$ and are used for event selection and \PbPb collision centrality
determination. Two beam scintillator counters (BSC) are used for triggering and
beam-halo rejection. A more detailed description of the CMS detector, together
with a definition of the coordinate system used and the relevant kinematic
variables, can be found in Ref.~\cite{Chatrchyan:2008zzk}.

The measurements reported here are based on \PbPb and \pp events triggered by a
hardware-based dimuon trigger without an explicit muon-momentum threshold.
Inelastic hadronic \PbPb collisions are selected using BSC and HF information,
in coincidence with a bunch crossing~\cite{Chatrchyan:2008zzk}. Events are
further filtered offline by requiring a reconstructed primary vertex and at
least three towers in each HF with an energy deposit of more than 3\GeV per
tower.
Muons are reconstructed offline using tracks in the muon detectors (``standalone
muons'') that are then matched to tracks in the silicon tracker, using an
algorithm optimized for the heavy-ion
environment~\cite{dEnterria:2007xr,Roland:2006kz}. In addition, an iterative
track reconstruction algorithm~\cite{Chatrchyan:2014fea} is applied to the \PbPb
data, limited to cone regions defined by the standalone muons. The \pp
reconstruction algorithm includes an iterative tracking step in the full silicon
tracker. The final parameters of the muon trajectory are obtained from a global
fit of the standalone muon with a track in the silicon tracker.
The single muon acceptance and identification criteria are the same as in
Ref.~\cite{Chatrchyan:2012np}. Opposite-sign muon pairs are fitted with a common
vertex constraint and are kept if the fit $\chi^2$ probability is greater
than 1\%.
Most of the non-prompt \JPsi and \psiP mesons, originating from \cPqb-hadron
decays, are rejected using the pseudo-proper decay length, $\ell_{\psi} =
L_{xy}\, m_{\psi}/\pt$, where $L_{xy}$ is the transverse distance between the
\mumu vertex and the interaction point and $m_{\psi}$ is the \JPsi or \psiP
mass. The $\ell_{\psi}$ selection condition is tuned with Monte Carlo (MC)
simulation studies, separately for the \pp and \PbPb collision systems, such
that 90\% of the prompt \JPsi and \psiP are kept, typically rejecting 80\% of
the non-prompt ones.
For these studies, unpolarized prompt and non-prompt \JPsi and \psiP mesons are
generated with \PYTHIA~6.424~\cite{Sjostrand:2006za} and decayed with
\EVTGEN~\cite{Lange:2001uf}, while the final-state bremsstrahlung is simulated
with \PHOTOS~\cite{Barberio:1993qi}. The signal events are embedded in
underlying heavy-ion events, generated with \HYDJET~1.8~\cite{Lokhtin:2005px},
at the level of detector hits and with matching vertices. The detector response
is simulated with \GEANTfour~\cite{Agostinelli:2002hh} and the resulting
information is processed through the full event reconstruction chain, including
trigger emulation.
The selection efficiency cancels in the double ratio and the remaining
non-prompt contamination is accounted for as a systematic uncertainty.

The analysis is performed in two dimuon kinematic domains: the ``midrapidity''
domain covers the range $\abs{y}<1.6$, where the \JPsi and \psiP mesons are only
reconstructed for $\pt>6.5\GeVc$, while the ``forward rapidity'' domain covers
the range $1.6<\abs{y}<2.4$, where the acceptance extends down to $\pt=3\GeVc$.
Dimuons are restricted to $\pt<30\GeVc$ in order to have a well defined
kinematic interval. The available \PbPb data at forward rapidity could not be fitted
reliably when split into the intervals $3<\pt<6.5\GeVc$ and $6.5<\pt<30\GeVc$.
Therefore, this analysis cannot differentiate between \pt and rapidity dependent
effects on the measured double ratios. The \PbPb sample is split in three bins
of collision centrality, defined using fractions of the inelastic hadronic cross
section where 0\% denotes the most central collisions: 40--100\%, 20--40\%, and
0--20\%. This fraction is determined from the HF energy
distribution~\cite{Chatrchyan:2011pb}. Related variables, such as the number of
nucleons participating in the collision (\npart), are evaluated using a
Glauber-model calculation~\cite{Chatrchyan:2011sx} and are only used to display
the centrality dependence of the measurements. The average \npart values
corresponding to the three centrality bins above are
$\langle\npart\rangle=32.8$, 158.7, and 308.4, respectively.

\begin{figure*}[t]
  \centering
  \includegraphics[width=0.48\textwidth]{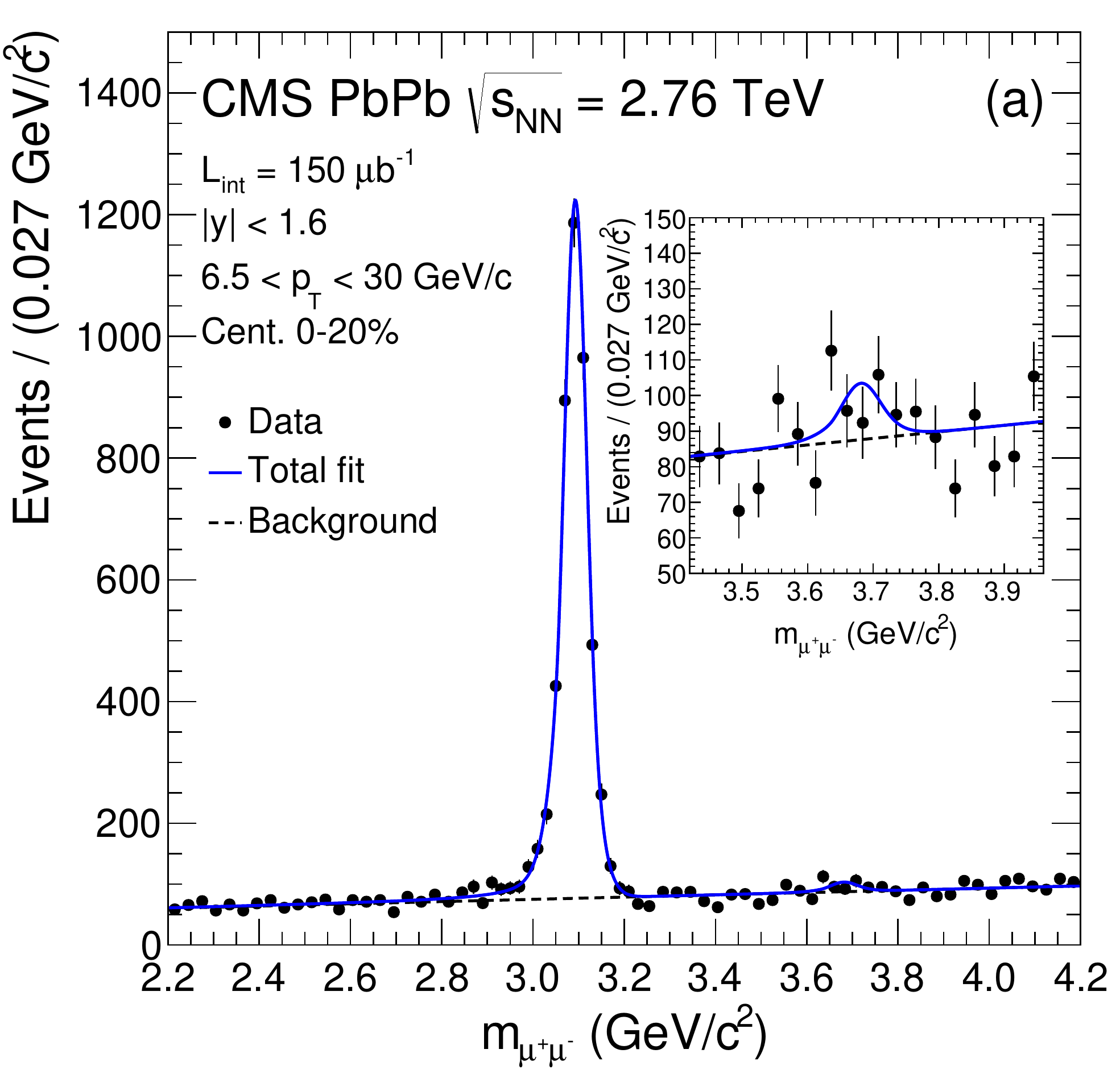}
  \includegraphics[width=0.48\textwidth]{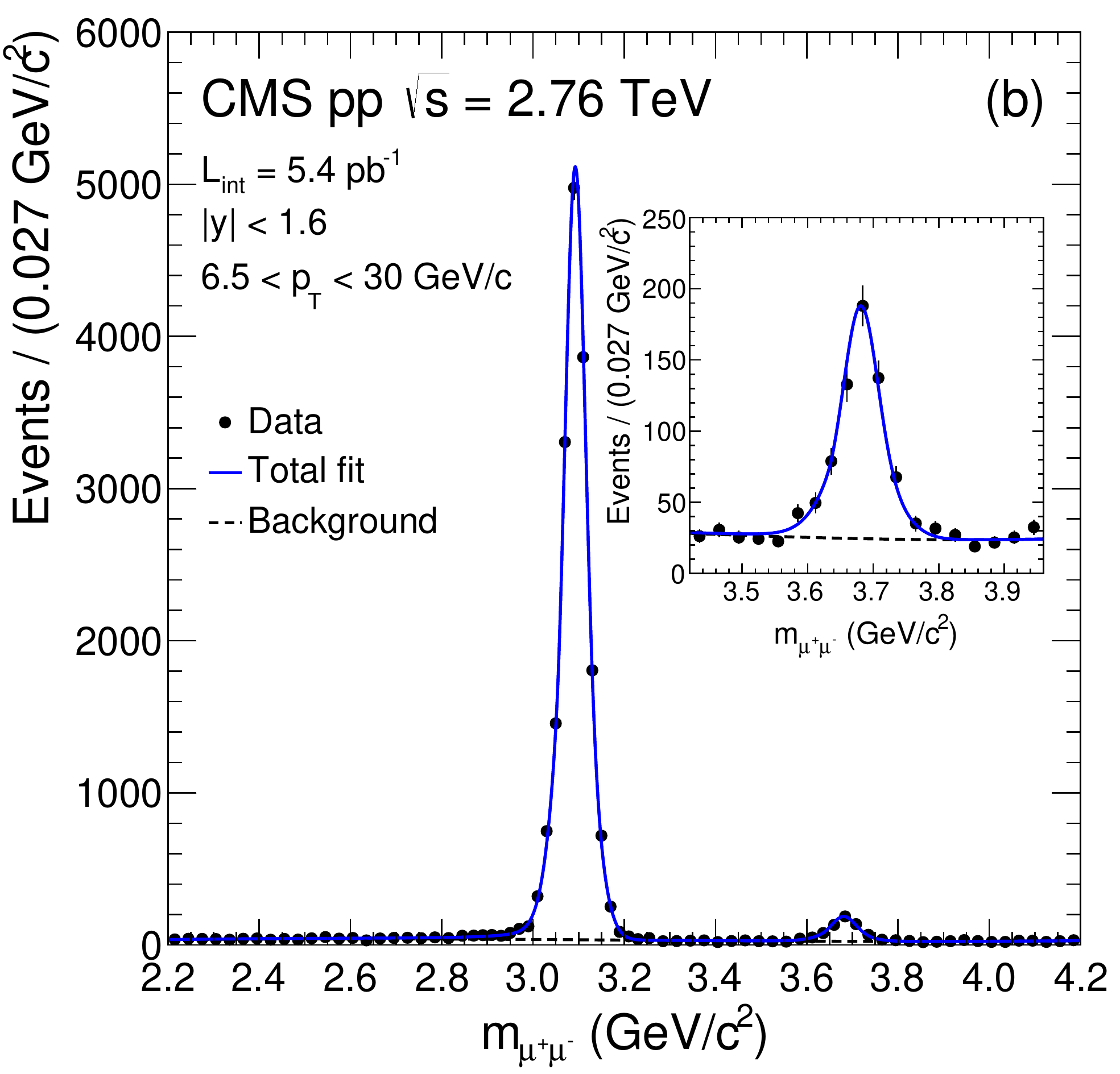}
  \includegraphics[width=0.48\textwidth]{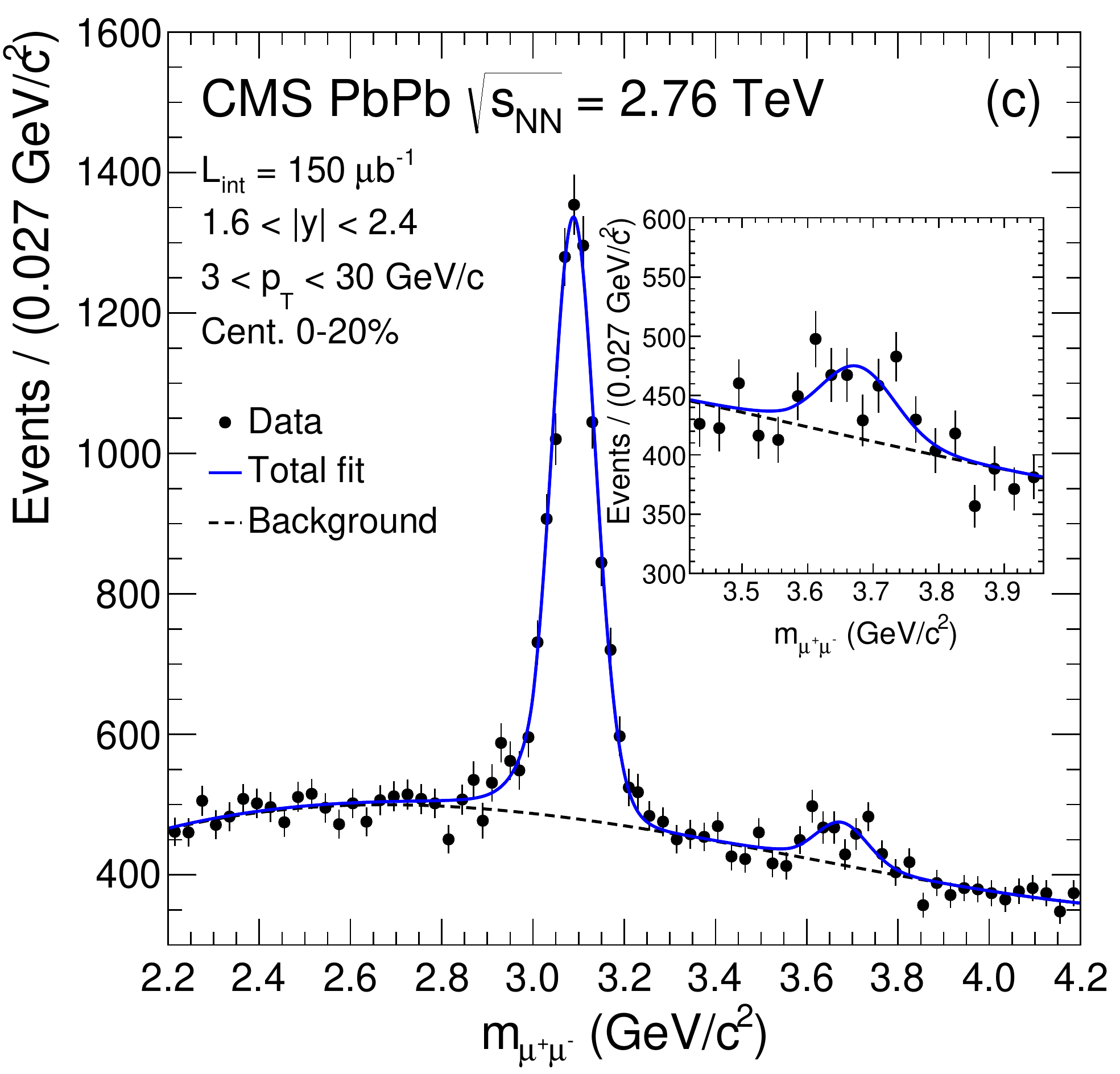}
  \includegraphics[width=0.48\textwidth]{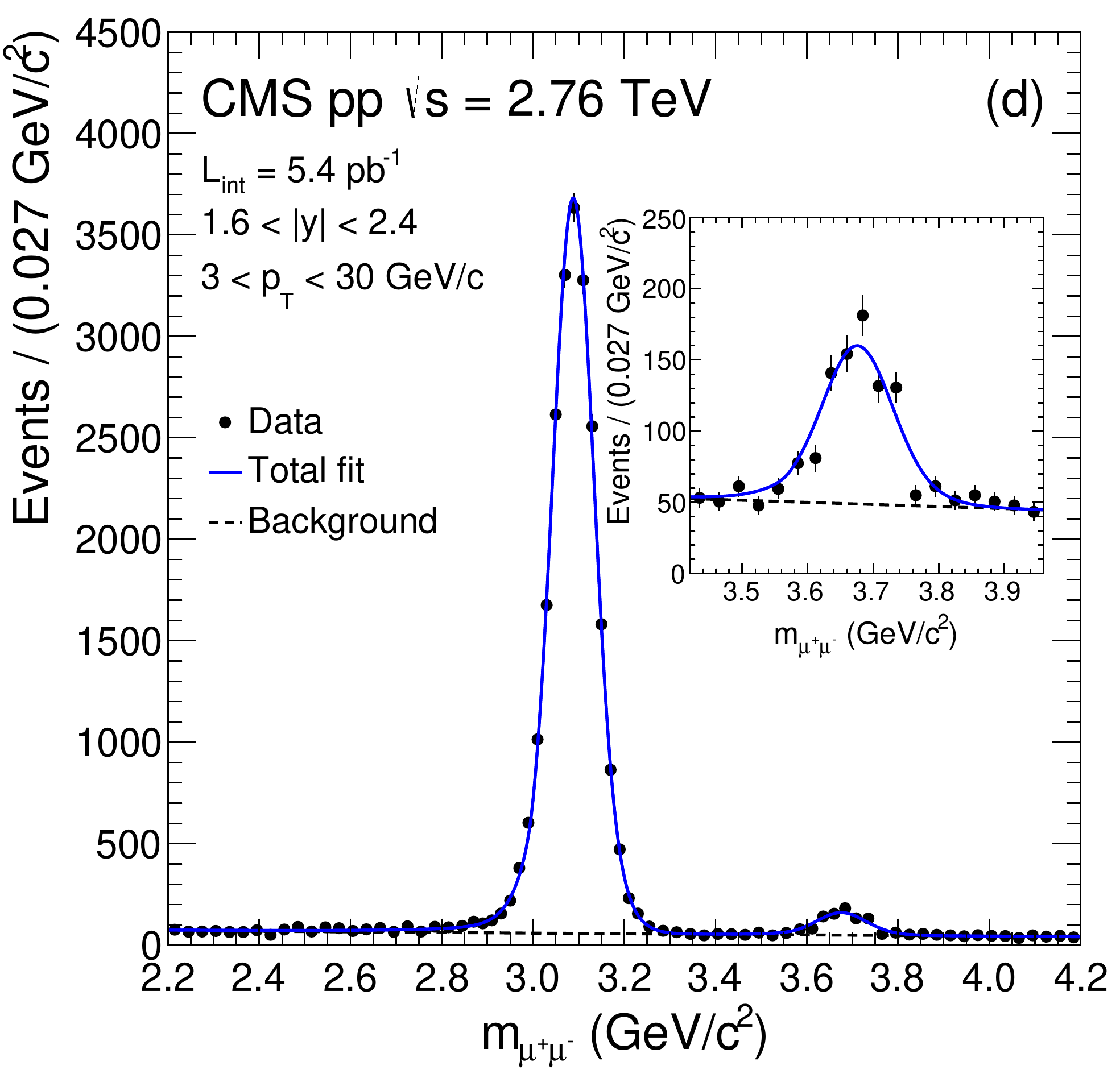}
  \caption{Dimuon invariant-mass distributions measured in central \PbPb (left)
    and \pp (right) collisions, for $\abs{y}<1.6$ and $6.5<\pt<30\GeVc$ (top) as
    well as $1.6<\abs{y}<2.4$ and $3<\pt<30\GeVc$ (bottom). The results of unbinned
    maximum likelihood fits are also shown. The \psiP region is magnified in the
    insets.}
  \label{fig:invmass}
\end{figure*}

Figure~\ref{fig:invmass} shows the dimuon invariant-mass ($m_{\mumu}$)
distributions measured in central (0--20\%) \PbPb and \pp collisions, for the
midrapidity and forward rapidity bins. The results of unbinned maximum
likelihood fits are also shown. Each charmonium resonance is described by the
sum of a Gaussian function and a Crystal Ball (CB) function~\cite{SLAC-R-236},
with common mean $m_0$, independent widths $\sigma_{\mathrm{G}}$ and
$\sigma_{\mathrm{CB}}$, and relative contribution of the Gaussian to the signal
yield $f_{\mathrm{G}}$.
In all cases, the fitted \JPsi mean agrees within 0.3\% with the world
average~\cite{Agashe:2014kda}. The resolution, after averaging the Gaussian and
CB widths, is about 30\MeVcc at midrapidity and 50\MeVcc at forward rapidity,
both for \pp and \PbPb data.
The CB radiative tail parameters $\alpha$ and $n$, common to both charmonia, are fixed to the
values obtained in fits to simulated distributions.
The $m_0$, $\sigma_{\mathrm{G}}$, and $\sigma_{\mathrm{CB}}$ parameters of the \psiP
resonance shape match the \JPsi parameters, scaled by the ratio of their masses,
$m_{\psiP}/m_{\JPsi}$~\cite{Agashe:2014kda}. This scaling assumption has been
validated in analyses of larger event
samples~\cite{Chatrchyan:2011kc,Chatrchyan:2012lxa}. The same value for
$f_{\mathrm{G}}$ is used in the definition of the \psiP and \JPsi signal shapes.
Six parameters are left free in the \pp fit: $m_0$, $\sigma_{\mathrm{G}}$,
$\sigma_{\mathrm{CB}}$, $f_{\mathrm{G}}$, the \JPsi yield, and the \psiP to \JPsi
yield ratio. In the \PbPb fits, instead, the double ratio replaces the \psiP to
\JPsi ratio as one of the free parameters. In addition, given their smaller
signal-to-background ratio, the \PbPb data are fitted fixing the
$\sigma_{\mathrm{G}}/\sigma_{\mathrm{CB}}$ ratio to the value obtained in fits to MC
distributions.

The background is described by Chebychev polynomials, of order ($0 \leq N \leq
3$) determined for each analysis bin with log-likelihood ratio (LLR) tests. The
background shape is mostly determined by the kinematic distributions of the
muons produced in meson decays, which are expected to change with collision
centrality~\cite{CMS:2012aa,ALICE:2012ab,Chatrchyan:2012np}.
Once the background functions are selected, the \pp and three \PbPb centrality
samples are fitted simultaneously. Since the signal shape does not depend on the
collision centrality~\cite{Chatrchyan:2012np}, the three \PbPb centrality bins
are fitted with common signal shape parameters, which are independent of the \pp
values; the four background shapes are independent. The simultaneous fit
directly provides the three double ratios (one per centrality class), for each
rapidity interval.

The systematic uncertainties from the fitting method are studied by varying the
signal and background shapes as well as the fitted invariant-mass range. As an
alternative signal shape, the sum of two CB functions with common mean and tail
parameters is used, leaving all parameters free in the fit except for the mass
scaling between the \JPsi and \psiP means and widths.
The uncertainty on the background is evaluated by considering three fit
variations: (i)~use as background shape an exponential function with a Chebychev
polynomial of order $1 \leq N \leq 3$ (determined with a LLR test) as an
argument; (ii)~extend the fitted mass region to $1.8<m_{\mumu}<5\GeVcc$;
(iii)~fit the \JPsi and \psiP regions (below 3.5\GeVcc and above 3.3\GeVcc,
respectively) with independent background functions. The maximum deviation from
the nominal fit is added in quadrature with the signal shape uncertainty to
obtain the fit systematic uncertainty in the double ratio, which varies between
8\% at midrapidity and 28\% at forward rapidity. The dominant contribution to
this uncertainty changes from bin to bin because of the strongly varying
signal-to-background ratio.
A cross-check made on the centrality-integrated sample shows that counting the
signal yields above a polynomial exclusively fitted to the sidebands gives a
result consistent with the nominal values.

Assuming no change in their polarizations, the \JPsi and \psiP acceptances are
independent of the collision system and cancel in the double ratio. Residual
effects from imperfect double-ratio cancellations of the muon reconstruction and
trigger efficiency corrections have been evaluated with MC simulation studies.
The MC double ratio of the signal efficiencies is compatible with unity. The MC
statistical uncertainties, 1\% at midrapidity and 5\% at forward rapidity, are
assigned as the systematic uncertainties on the assumption that the efficiency
corrections cancel. Differences between the MC and data
distributions~\cite{Chatrchyan:2012np} have a negligible impact on the
efficiency double ratio.
The same charmonium \pt and rapidity distributions have been used in the
generation of \pp and \PbPb events. The efficiency double-ratio varies by
less than 1\% when changing the kinematic
distributions in \PbPb within a reasonable range, evaluated using \raa
measurements as a function of \pt and rapidity~\cite{Chatrchyan:2012np}.

The prompt charmonia are simulated unpolarized, a scenario in good agreement
with \pp measurements~\cite{Abelev:2011md,Chatrchyan:2013cla,Aaij:2013nlm}.
Alternative scenarios have been considered, where the polarizations change
significantly from the \JPsi to the \psiP and/or from \pp to \PbPb collisions.
The impact is completely negligible on the double ratio of the measurement
efficiencies, while the double ratio of the acceptances can change by
${\approx}20\%$. As in previous
analyses~\cite{Chatrchyan:2011pe,Chatrchyan:2012lxa,Chatrchyan:2013nza}, such
possible physics effects are not considered as systematic uncertainties.

In \pp collisions, around 20--25\% of the charmonium yields are due to
\cPqb-hadron decays~\cite{Chatrchyan:2011kc}; no evidence for different values
has been seen in \PbPb collisions~\cite{Chatrchyan:2012np}. Considering a
\cPqb-hadron rejection inefficiency of about 20\%, the prompt \JPsi and \psiP
yields include a residual contamination from \cPqb-hadron decays of up to 5\%.
These \cPqb-hadron contaminations are conservatively assumed to be independent
in the four yields entering the double ratio.

Adding in quadrature the uncertainties mentioned above leads to total systematic
uncertainties of 13--30\%, values smaller than the corresponding statistical
uncertainties.

\begin{figure}[ht]
\centering
\includegraphics[width=\cmsFigWidth]{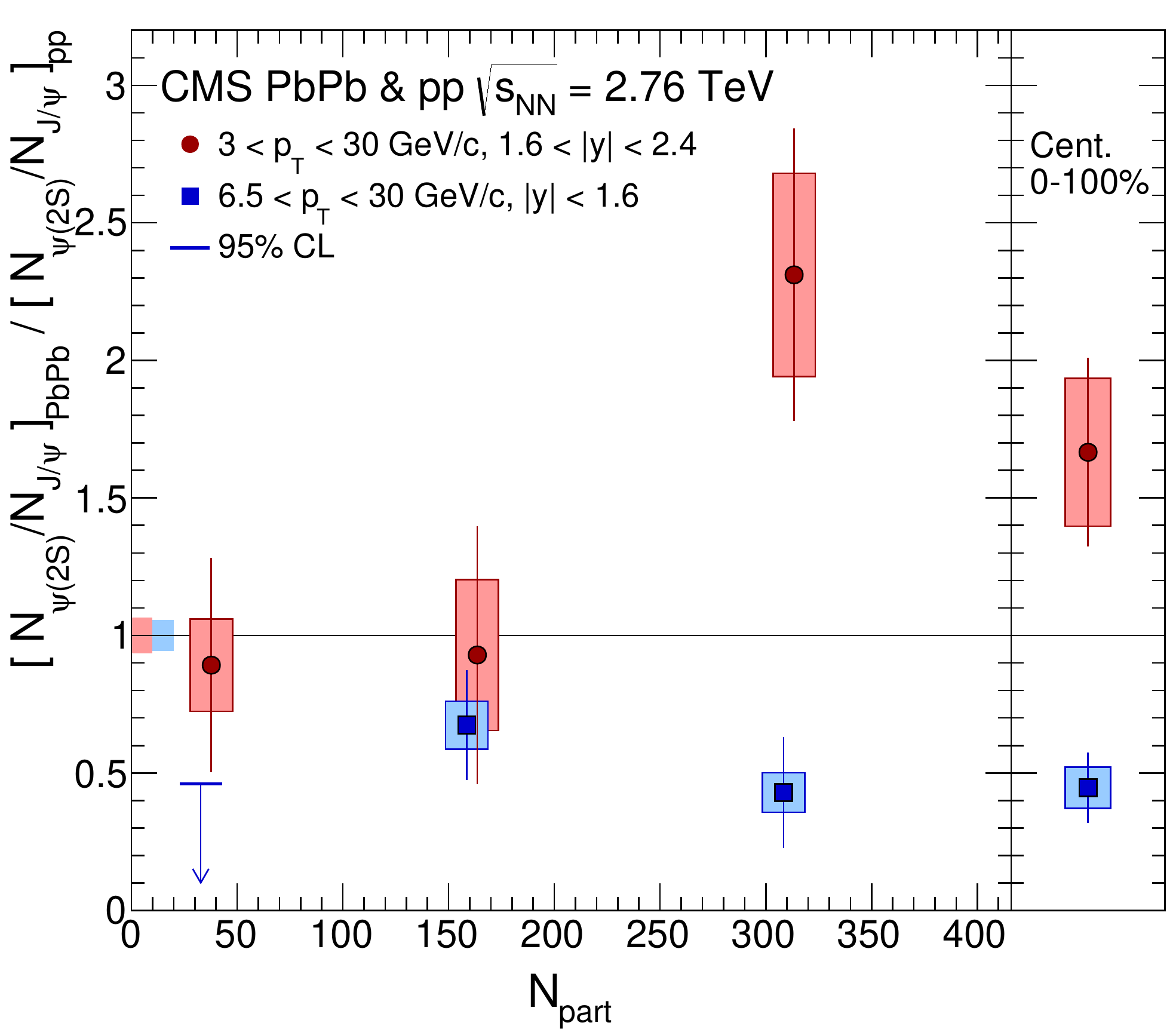}
\caption{Double ratio of measured yields, \doubleRatio, as a function of
  centrality, for the midrapidity (blue squares) and forward rapidity (red
  circles, slightly shifted) analysis bins. The centrality-integrated results
  are displayed in the right panel. Statistical (systematic) uncertainties are
  shown as bars (boxes). The boxes at unity indicate the (global) \pp
  uncertainties.}
\label{fig:doubleratio_cent}
\end{figure}

The double ratio of measured yields, $(N_{\psiP} / N_{\JPsi})_{\PbPb} /
(N_{\psiP} / N_{\JPsi})_{\pp}$, is shown in
Fig.~\ref{fig:doubleratio_cent} as a function of centrality, for both kinematic
bins.
The quadratic sum of the \pp statistical and systematic uncertainties
(${\approx}6\%$) is common to all centralities. The centrality-integrated
results are also displayed, in the right panel. In the most peripheral \PbPb
collisions, no significant \psiP signal has been observed in the midrapidity bin
and an upper limit of 0.47 at 95\% confidence level (\CL) is set on the double
ratio, using the Feldman--Cousins method~\cite{Feldman:1997qc}.

In the midrapidity bin, restricted to $\pt>6.5\GeVc$,
the double ratio is below unity in all centrality bins, with a
centrality-integrated value of
$0.45\pm0.13\stat\pm0.07\syst$, including the global \pp
uncertainties.
In the forward rapidity bin, which extends down to $\pt=3\GeVc$, the
centrality-integrated double ratio increases to
$1.67\pm0.34\stat\pm0.27\syst$. While the forward-rapidity
double ratio is consistent with unity in peripheral \PbPb collisions, it becomes
$2.31\pm0.53\stat\pm0.37\syst \pm0.15\,\text{(pp)}$ in the
20\% most central collisions, indicating that the \psiP to \JPsi yield ratio is
enhanced in central \PbPb collisions with respect to \pp collisions (the
hypothesis of being compatible with unity has a p-value of only 0.011).

Nuclear modification factors for prompt \psiP production, $\raa(\psiP)$, can be
derived by multiplying the centrality-integrated double ratios by the
corresponding prompt \JPsi \raa which can be found in
Ref.~\cite{Chatrchyan:2012np}. The resulting centrality-integrated \raa values
for \psiP are $0.13 \pm 0.04\stat \pm 0.02\syst \pm
0.01\,(\pp)$ at midrapidity and $0.67 \pm 0.16\stat \pm
0.11\syst \pm 0.07\,(\pp)$ at forward rapidity. While the
$\psiP/\JPsi$ ratio at low \pt (forward rapidity) is enhanced in central \PbPb
collisions, as compared to \pp, the yield of \psiP itself in \PbPb collisions is
still suppressed in comparison to the yield in \pp collisions scaled by the
number of inelastic nucleon-nucleon collisions.

In summary, the CMS measurements reported in this Letter show two interesting
observations. First, \psiP production is suppressed in \PbPb collisions with
respect to \pp collisions, in both kinematic regions investigated. Second, in
comparison to \JPsi production and in the most central \PbPb collisions, \psiP
production is suppressed in the range $\abs{y}<1.6$ and $6.5<\pt<30\GeVc$, as
expected in the sequential melting scenario and matching the corresponding
bottomonia pattern~\cite{Chatrchyan:2012lxa}, while it is enhanced in the range
$1.6<\abs{y}<2.4$ and $3<\pt<30\GeVc$. Such behavior implies the presence of
physics processes that either cause the \pt dependence of $\raa(\psiP)$ to be
weaker than for the $\raa(\JPsi)$ or cause the $\raa(\psiP)$ to start decreasing
at higher \pt. Alternatively, these processes would have to have the opposite
dependence with increasing rapidity. Larger event samples are needed to evaluate
in more detail how these observations depend separately on the \pt and rapidity
of the charmonium states.

\begin{acknowledgments}
We congratulate our colleagues in the CERN accelerator departments for the excellent performance of the LHC and thank the technical and administrative staffs at CERN and at other CMS institutes for their contributions to the success of the CMS effort. In addition, we gratefully acknowledge the computing centers and personnel of the Worldwide LHC Computing Grid for delivering so effectively the computing infrastructure essential to our analyses. Finally, we acknowledge the enduring support for the construction and operation of the LHC and the CMS detector provided by the following funding agencies: BMWFW and FWF (Austria); FNRS and FWO (Belgium); CNPq, CAPES, FAPERJ, and FAPESP (Brazil); MES (Bulgaria); CERN; CAS, MoST, and NSFC (China); COLCIENCIAS (Colombia); MSES and CSF (Croatia); RPF (Cyprus); MoER, ERC IUT and ERDF (Estonia); Academy of Finland, MEC, and HIP (Finland); CEA and CNRS/IN2P3 (France); BMBF, DFG, and HGF (Germany); GSRT (Greece); OTKA and NIH (Hungary); DAE and DST (India); IPM (Iran); SFI (Ireland); INFN (Italy); NRF and WCU (Republic of Korea); LAS (Lithuania); MOE and UM (Malaysia); CINVESTAV, CONACYT, SEP, and UASLP-FAI (Mexico); MBIE (New Zealand); PAEC (Pakistan); MSHE and NSC (Poland); FCT (Portugal); JINR (Dubna); MON, RosAtom, RAS and RFBR (Russia); MESTD (Serbia); SEIDI and CPAN (Spain); Swiss Funding Agencies (Switzerland); MST (Taipei); ThEPCenter, IPST, STAR and NSTDA (Thailand); TUBITAK and TAEK (Turkey); NASU and SFFR (Ukraine); STFC (United Kingdom); DOE and NSF (USA).
\end{acknowledgments}

\bibliography{auto_generated}  % will be created by the tdr script.
\cleardoublepage \appendix\section{The CMS Collaboration \label{app:collab}}\begin{sloppypar}\hyphenpenalty=5000\widowpenalty=500\clubpenalty=5000\input{HIN-12-007-authorlist.tex}\end{sloppypar}
\end{document}

%% file: HIN-12-007-authorlist.tex
\textbf{Yerevan Physics Institute,  Yerevan,  Armenia}\\*[0pt]
V.~Khachatryan, A.M.~Sirunyan, A.~Tumasyan
\vskip\cmsinstskip
\textbf{Institut f\"{u}r Hochenergiephysik der OeAW,  Wien,  Austria}\\*[0pt]
W.~Adam, T.~Bergauer, M.~Dragicevic, J.~Er\"{o}, C.~Fabjan\cmsAuthorMark{1}, M.~Friedl, R.~Fr\"{u}hwirth\cmsAuthorMark{1}, V.M.~Ghete, C.~Hartl, N.~H\"{o}rmann, J.~Hrubec, M.~Jeitler\cmsAuthorMark{1}, W.~Kiesenhofer, V.~Kn\"{u}nz, M.~Krammer\cmsAuthorMark{1}, I.~Kr\"{a}tschmer, D.~Liko, I.~Mikulec, D.~Rabady\cmsAuthorMark{2}, B.~Rahbaran, H.~Rohringer, R.~Sch\"{o}fbeck, J.~Strauss, A.~Taurok, W.~Treberer-Treberspurg, W.~Waltenberger, C.-E.~Wulz\cmsAuthorMark{1}
% ...
\vskip\cmsinstskip
% ...
1:~~Also at Vienna University of Technology, Vienna, Austria\\
2:~~Also at CERN, European Organization for Nuclear Research, Geneva, Switzerland\\
% ...
